\documentclass[a4paper]{spie} 
\usepackage{amsmath,amsfonts,amssymb}
\usepackage{graphicx}
\usepackage{setspace}
\usepackage{tocloft}
\usepackage{graphicx}
\usepackage{epstopdf}
\usepackage{subcaption}
\usepackage{empheq}
\usepackage{url,xurl}
\usepackage[colorlinks=true, allcolors=blue]{hyperref}

\usepackage{float}
\usepackage[skip=1em , belowskip=4pt]{caption}
\usepackage{subcaption}
\usepackage{booktabs}
\usepackage{tikz}
\usetikzlibrary{angles,quotes}
\usetikzlibrary{quantikz}
\usepackage[most]{tcolorbox}

\title{Simulation and analysis of quantum phase estimation algorithm in the presence of incoherent quantum noise channels}
\author{Muhammad Faizan and Muhammad Faryad$^\ast$
\skiplinehalf
Lahore University of Management Sciences, Department of Physics,  Lahore 54792, Pakistan
}

\authorinfo{$^\ast$Corresponding author: \href{mailto:muhammad.faryad@lums.edu.pk}{muhammad.faryad@lums.edu.pk}}

\begin{document}
\maketitle

\begin{abstract}
The quantum phase estimation (QPE) is one of the fundamental algorithms based on the quantum Fourier transform. It has applications in order-finding, factoring, and finding the eigenvalues of unitary operators. The major challenge in running QPE and other quantum algorithms is the noise in quantum computers. In the present work, we study the impact of incoherent noise on QPE, modeled as trace-preserving and completely positive quantum channels. Different noise models such as depolarizing, phase flip, bit flip, and bit-phase flip are taken to understand the performance of the QPE in the presence of noise. The simulation results indicate that the standard deviation of the eigenvalue of the unitary operator has strong exponential dependence upon the error probability of individual qubits. However, the standard deviation increases only linearly with the number of qubits for fixed error probability when that error probability is small.
\end{abstract}

\section{Introduction}
The major stumbling block in realizing the full potential of quantum computers is the noise in the currently available quantum hardware. This noise is due to the interactions of qubits with the environment and due to the faulty gate operations \cite{nielsen2002quantum}.  The noise in the qubits can be classified as coherent or incoherent. The coherent noise is primarily due to the miscalibration of pulses implementing various gate sequences. In contrast, the incoherent noise arises due to the interaction of qubits with the environment and other qubits. It gives rise to the qubit’s decoherence and limits the available time to implement quantum algorithms, thus producing statistical noise in the output \cite{preskill2018quantum}. Therefore, understanding incoherent noise and its impact on the performance of quantum algorithms is important in evaluating the performance of various quantum algorithms. In this work, we analyze the impact of unital incoherent noise processes on the quantum phase estimation (QPE) algorithm.
QPE is one of the most important subroutines in quantum computation, based on quantum Fourier transform, and serves as a building block for many other algorithms, including Shor's Algorithm \cite{shor1994algorithms}.

Various studies investigating the performance of QPE in the noisy environment have been conducted. The performance of QPE has been studied in the presence of static gate defect \cite{garcia2008quantum}, residual coupling between qubits, decoherence \cite{barenco1996approximate}, and coherent phase errors \cite{wei2004quantum}. The robustness of QPE has been investigated while implementing it physically to simulate the molecular energy in the presence of noise and decoherence \cite{paesani2017experimental}. The performance of QPE in the context of fundamental noise channels was recently explored \cite{Abdullah}; however, we extend that work and investigate the effect of noise on the standard deviation of the output in the present work. 

 In the presence of noise, the evolution of a qubit is no longer unitary. Instead, its evolution can be captured using completely positive and trace-preserving maps of the density operators \cite{fano1957description,choi1975completely}, commonly known as quantum channels.  Mathematically, quantum channels can be decomposed using various representations and can be viewed as modeling the interaction of a quantum system with an environment \cite{kraus1983states,stinespring1955positive}. In this paper, we will study the impact of (1) bit flip, (2) phase flip, (3) bit-phase flip, and (4) depolarizing channels on QPE. These four noise processes are fundamental in the sense that other single-qubit unital noise processes can be described in terms of these processes. This paper is organized as follows: In Section \ref{methods}, we provide a brief overview of the QPE algorithm, noise models, and incorporation of noise models in the QPE algorithm. The simulation results and discussion is presented in Section \ref{srd}. The conclusions are presented in Section \ref{conclusion}.

\section{Modeling Noise in QPE}\label{methods}
\begin{figure}[H]
	\centering
	\scalebox{1.2}{\begin{quantikz}
			\lstick[wires=4]{$\ket{0}^{\otimes n}$} &[2mm] \gate{H}
			& \ctrl{4} & \qw & \qw & \qw & \gate[wires=4, nwires={3}][2cm]{FT^\dagger} & \meter{} & \rstick[wires=4]{$\ket{2^n\theta}$}\qw \\
			& \gate{H} & \qw & \ctrl{3} & \qw & \qw & & \meter{} & \qw \\
			& \vdots & & &\vdots & & & \vdots & \\
			& \gate{H} & \qw & \qw & \qw & \ctrl{1} & & \meter{} & \qw  \\
			\lstick{$\ket{v}$}
			& \qw & \gate{U} & \gate{U^{2}} &\qw  \cdots &  \gate{U^{2^{n-1}}}& \qw & \qw  & \qw \rstick{$\ket{v}$} 
	\end{quantikz}}

	\caption{The circuit for quantum phase estimation for estimating the eigenvalue of a unitary operator $U$.}
	\label{QPECircuit}
\end{figure}
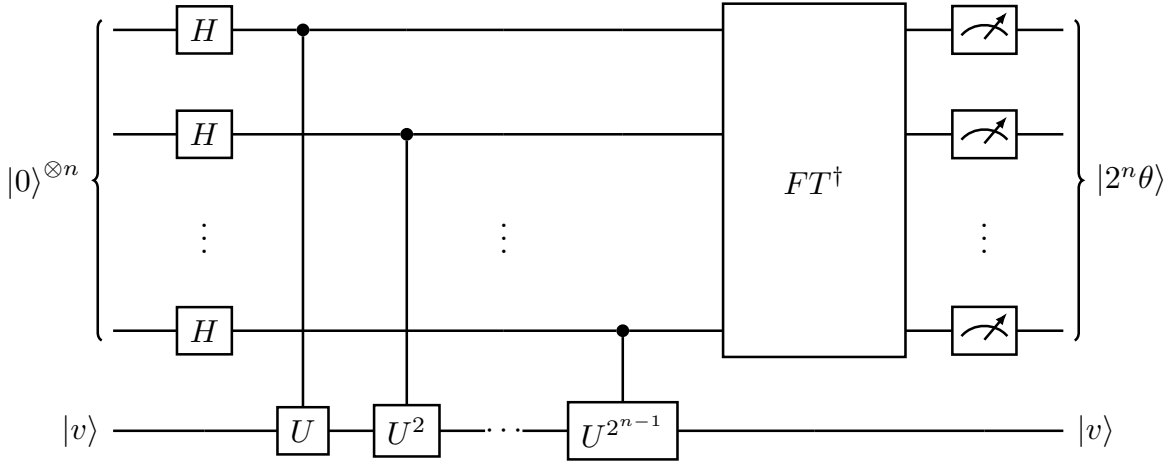

The QPE  algorithm  finds the eigenvalue of a unitary operator $U$ such that $U\ket{v}=e^{2\pi i\theta}\ket{v}$ where $0\leq\theta\leq 1$ and $\ket{v}$ is an eigenvector of $U$. The main idea of this algorithm is to provide an $n$-bit approximation of $\theta$ in a single run. For the work presented in this paper, we took $U$ as a single qubit operator that has eigenvectors $\ket{0}$ and $\ket{1}$ with eigenvalues $1$ and $e^{2\pi i\theta}$, respectively.
This algorithm uses two registers. The first register is initialized to $\ket{0}^{\otimes n}$, where $n$ is the number of qubits to estimate $\theta$. The second register starts in the state $\ket{v}$ and contains the number of qubits necessary to store $\ket{v}$. The algorithm is performed in four steps: (1) creating superposition of all possible states and performing the controlled-$U^{2^j}$ operation for $j\geq 0$, (2) applying the \emph{inverse} quantum Fourier transform on the first register, (3) measurement in the computational basis to read out the state of the first register, and (4) classical post-processing to extract estimated $\theta$. The circuit for quantum phase estimation is illustrated in Fig. (\ref{QPECircuit}) for the simplest unitary operator $U$ acting on a single qubit \cite{nielsen2002quantum}.

\begin{figure}[htbp]
	\centering
	\subfloat{{\includegraphics[width=0.4\textwidth]{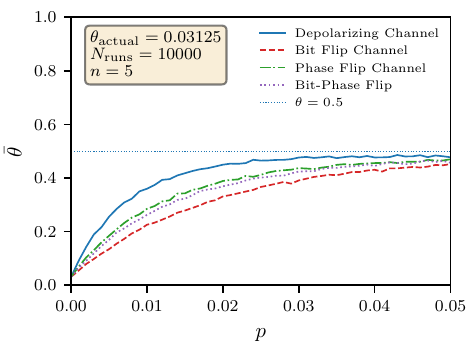} }}
	\subfloat{{\includegraphics[width=0.4\textwidth]{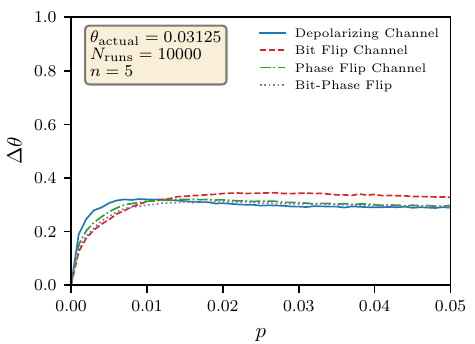} }}\\
	\subfloat{{\includegraphics[width=0.4\textwidth]{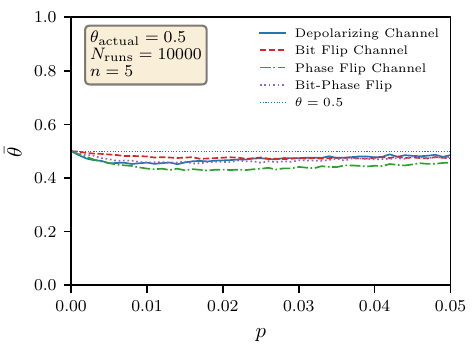} }}
	\subfloat{{\includegraphics[width=0.4\textwidth]{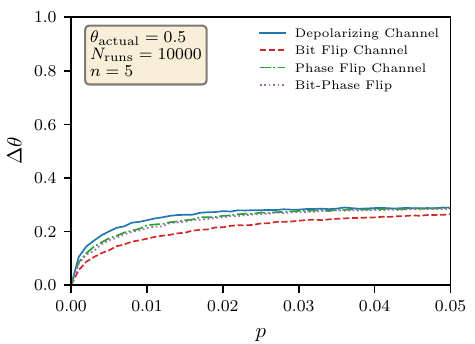} }}\\
	\subfloat{{\includegraphics[width=0.4\textwidth]{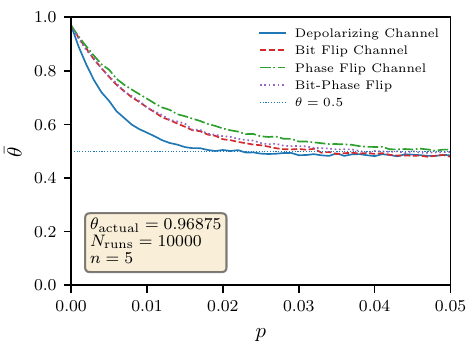} }}
	\subfloat{{\includegraphics[width=0.4\textwidth]{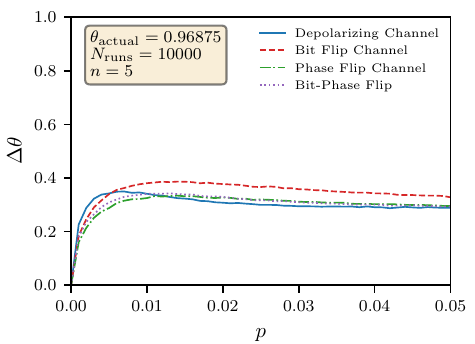} }}
	\caption{Average value $\overline{\theta}$ and standard deviation $\Delta\theta$ plotted as a function of error probability $p$ for three different values of the actual $\theta$ when QPE is implemented with $n=5$ qubits. A dotted line at $\theta=0.5$ is added for reference in the left panels.}
	\label{qpeplots5}
\end{figure}

For the simulation results in this paper, we considered four noisy quantum channels to represent noise processes. The first is the bit flip channel that represents a noise process that flips the qubit with probability $p$ and is given as$$\mathcal{E}(\rho)=(1-p)\rho + p X\rho X,$$ where $\rho$ is the density matrix representing the state of a single qubit and $X$ is the Pauli $X$ gate. This channel applies the $X$ gate to the qubit with probability $p$.

The second noise model we used in this work is the phase-flip channel that models the random flipping of the phase of the qubit state $\ket{1}$. This noise model can be written as $$\mathcal{E}(\rho)=(1-p)\rho+pZ\rho Z,$$where $Z$ is the Pauli $Z$ gate.

The third noise model we considered is the combination of bit-flip and phase flip channel, called bit-phase flip channel defined as$$\mathcal{E}(\rho)=(1-p)\rho+pY\rho Y.$$ Since $Y=iXZ$, this channel applies both the phase flip and bit flip with probability $p$.

The fourth channel is called the depolarizing channel. Under this noise model,  the environment interacts with the system in such a way that with the probability $p$, a qubit is depolarized, i.e., the state of the system is transformed into a maximally mixed state, $I/2$. This channel is described by the map $$\mathcal{E}(\rho)=(1-p)\rho+p{I}/{2},$$where $I$ is the identity matrix describing a maximally mixed state of a single qubit.

To investigate the performance of QPE in the presence of noise, we transpiled the circuit to the one containing basic gate set $\{I , X , \sqrt{X} , R_z , CX\}$ and then incorporated 
the noise models into the circuit such that all the basic gates are followed by noise channel. To do this, we used qiskit's inbuilt noise models \cite{NoiseMod21:online}. 


\section{Simulation Results and Discussion}\label{srd}
To estimate the impact of four different noise processes on the eigenvalue of the unitary operator, we ran the QPE with noise for three different actual values $\theta_{\text{actual}}$ as a function of error probability $p$ by fixing $\ket{v}=\ket{1}$. We computed the average value of $\theta$ as $\overline{\theta}$ by running the algorithm $N_{\text{runs}}$ number of times for each value of $p$. We also computed the standard deviation $\Delta \theta$. The results for four noise models are presented in Fig. \ref{qpeplots5} for $n=5$ qubits to estimate the value of $\theta$. All the figures indicate that the average value of $\theta$ approaches $0.5$ as the error probability increases. Also, the standard deviation of the results increases with $p$ and then saturates as illustrated in Fig. \ref{qpeplots5}. This is because all four noise channels ultimately derive the output state of the QPE circuit (before measurement) to the maximally mixed state. The maximally mixed state gives all possible measurement results in an equal probability making  $\overline{\theta}$ approach $0.5$, which is the mean of all possible values of $\theta=\{0,1,2,..2^n-1\}/2^n$.  As $p$ increases, the output approaches the maximally mixed state. Furthermore, we observe that all channels have similar dependence on $p$ though the dependence is stronger for the depolarization channel than the other three channels because it combines all other channels into one.

\begin{figure}[htbp]
	\centering
	\begin{minipage}{0.4\linewidth}
		\centering
		\includegraphics[width=1\textwidth]{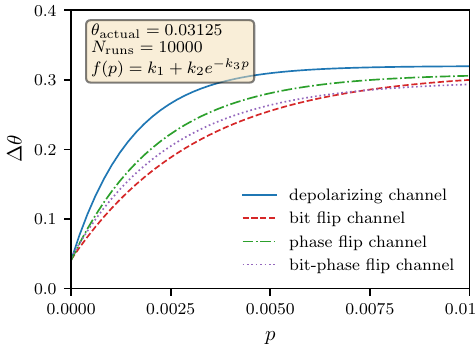}
	\end{minipage}\begin{minipage}{0.4\linewidth}
		\centering
		\includegraphics[width=1\textwidth]{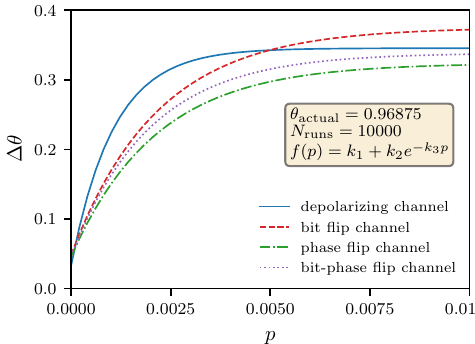}
	\end{minipage}
	\caption{Best fit plot of $\Delta \theta$ vs. $p$ for $n=5$ qubits. The model paramters are provided in Table \ref{tabb}.}
	\label{stdvsp}
\end{figure}

To characterize the approach of the circuit output towards the maximally mixed state, we computed the standard deviation $\Delta\theta$ of the measured output and presented them in Fig. \ref{qpeplots5}. The figure shows that $\Delta\theta$  starts small and approaches a saturation value when $p$ increases. This gives us a better measure of the approach to maximally mixed state even when the actual $\theta$ can be close to $0.5$. Since the results in Fig. \ref{qpeplots5} for $\Delta\theta$ show an exponential increase, we fitted $\Delta\theta$ to the following model, when $p$ is small,
\begin{equation}
	\Delta\theta(p) = k_1 + k_2e^{-k_3p}\,,\quad\,0\leq p\leq0.01\,,
\end{equation}
for all four channels and obtained parameters of the curve that best models the data with more than $98\%$ accuracy. The parameters are provided in Table \ref{tabb}. The table shows that $\Delta\theta$ has strong exponential dependence upon $p$. The plot of standard deviation $\Delta\theta$ vs. noise level $p$ such that $0<p<0.01$ is illustrated in Fig. \ref{stdvsp}. The results in Table \ref{tabb} indicate that the depolarization channel has the strongest exponential dependence on noise. Furthermore, $\Delta\theta$ approaches $k_1\approx0.3$ when actual $\theta$ is either close to $0$ or $1$ but $k_1\approx0.2$ when $\theta$ was $0.5$ when $p=0.01$. However, when $p=0.05$, $\Delta\theta$ approaches $0.3$ for all values of $\theta$ as can be seen from Fig. \ref{qpeplots5}. This can also be expected from the fact that the standard deviation of uniformly distributed data from $0$ to $1$ is about $0.3$ because a maximally mixed state at the output will lead to a uniform distribution of all possible values of $\theta\in\{0,1\}$.

\begin{table}[H]
	\centering
	\caption{The optimal set of parameters for the function $\Delta\theta(p) = k_1 + k_2e^{-k_3p}$ that best fits the standard deviation $\Delta\theta$ for $0\leq p\leq 0.01$ with more than $98\%$ accuracy.}
		\begin{subtable}[c]{0.30\textwidth}
		\centering
		\subcaption{$\theta_{\text{actual}}=0.03125$}
		\begin{tabular}[t]{lcccc}
			\toprule
			Noise Channel  & $k_1$ & $k_2$ & $k_3$\\
			\midrule
			Depolarizing    & 0.32 & -0.28 & 654 \\
			Bit flip  	    & 0.31 & -0.27 & 317 \\
			Phase flip      & 0.31 & -0.27 & 448 \\
			Bit-phase flip  & 0.30 & -0.26 & 413 \\
			\bottomrule
		\end{tabular}
	\end{subtable}\hfill 
	\begin{subtable}[c]{0.25\textwidth}
		\centering
		\subcaption{$\theta_{\text{actual}}=0.5$}
		\begin{tabular}[t]{lcccc}
			\toprule
			$k_1$ & $k_2$ & $k_3$\\
			\midrule
			0.24 & -0.21 & 338 \\
			0.18 & -0.16 & 238 \\
			0.22 & -0.19 & 293 \\
			0.22 & -0.19 & 267 \\
			\bottomrule
		\end{tabular}
	\end{subtable}	\hfill 
	\begin{subtable}[c]{0.25\textwidth}
		\centering
		\subcaption{$\theta_{\text{actual}}=0.96875$}
		\begin{tabular}[t]{lcccc}
			\toprule
			$k_1$ & $k_2$ & $k_3$\\
			\midrule
			0.35 & -0.31 & 927 \\
			0.38 & -0.33 & 460 \\
			0.33 & -0.28 & 448 \\
			0.34 & -0.29 & 508 \\
			\bottomrule
		\end{tabular}
		
	\end{subtable}
	\medskip

	\label{tabb}
\end{table}

\begin{figure}[htbp]
	\centering
	\subfloat{{\includegraphics[width=0.4\textwidth]{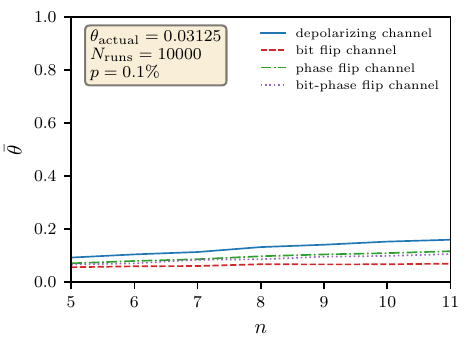} }}\subfloat{{\includegraphics[width=0.4\textwidth]{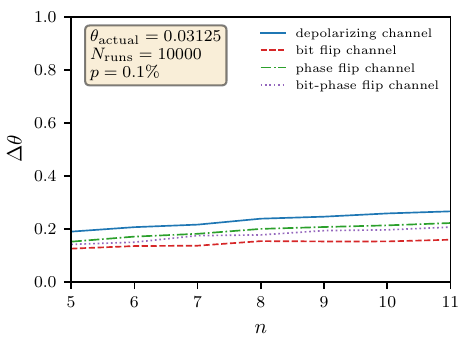} }}\\
	\subfloat{{\includegraphics[width=0.4\textwidth]{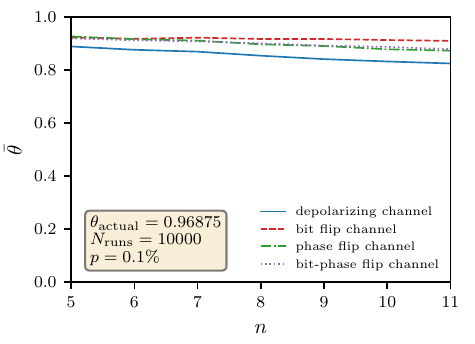} }}\subfloat{{\includegraphics[width=0.4\textwidth]{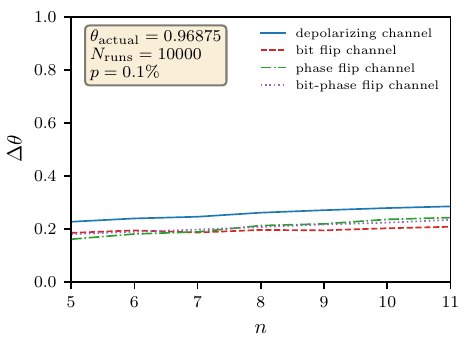} }}
	\caption{Average value  $\overline{\theta}$ and standard deviation $\Delta{\theta}$ as function of number of qubits $n$ for fixed $p$ and for two different values of actual $\theta$. }
	\label{fixedp}
\end{figure}

To understand the dependence of QPE output on the number of qubits $n$ for a fixed value of error probability $p$, we computed the average value of $\theta$ as a function of $n$ for two actual values of $\theta$. The results are plotted in Fig. \ref{fixedp}. The figure indicates that the increase in the number of qubits $n$ makes the output of the circuit diverge from the actual value of $\theta$ even though the noise level $p$ is fixed. Therefore, the increase in the number of qubits has an adverse impact on the results of the QPE algorithm as opposed to the expectation from a noiseless quantum computer where an increase in the number of qubits is expected to result in a better estimate of the $\theta$. However, the dependence of $\overline{\theta}$ and $\Delta\theta$ on $n$ is  only linear for small values of $p$ as opposed to very strong dependence on $p$.

\color{black}
\section{Conclusions}\label{conclusion}
We simulated the quantum phase estimation (QPE) algorithm with four noise processes: bit flip, phase flip, bit-phase flip, and depolarizing. We transpiled the algorithm into $ \{I , X , \sqrt{X} , R_z , CX\}$ gate set and introduced the noise channel after every gate with error probability $p$ for each qubit. The simulation results indicated that the average value of the eigenvalue of the unitary operator converged to $0.5$ regardless of the actual value of the eigenvalue $0\leq\theta\leq1$ as the error probability increased from $0$. This is because the noise processes force the overall quantum state towards the maximally mixed state. The standard deviation of the output increased exponentially for a small value of $p$ as $p$ increased from $0$ and then saturated. Furthermore, the average value of the eigenvalue diverged away from the actual value when the number of qubits was increased for fixed error probability, contrary to the noiseless QPE where increasing the number of qubits increases the precision of the eigenvalue. However, the dependence of the output on $n$ was linear as opposed to exponential dependence upon $p$.

\end{document}